# An Energy-efficient Clock Synchronization Protocol for Wireless Sensor Networks

Roxana Albu, Yann Labit, Thierry Gayraud, Pascal Berthou
CNRS ; LAAS ; 7, avenue du Colonel Roche, F-31077 Toulouse, France
Université de Toulouse ; UPS ; INSA, INP, ISAE ; LAAS ; F-31077 Toulouse, France
{ralbu, ylabit, berthou, gayraud}@laas.fr

Abstract—The behavior of Wireless Sensor Networks (WSN) is nowadays widely analyzed. One of the most important issues is related to their energy consumption, as this has a major impact on the network lifetime. Another important application requirement is to ensure data sensing synchronization, which leads to additional energy consumption as a high number of messages is sent and received at each node. Our proposal consists in implementing a combined synchronization protocol based on the IEEE 1588 standard that was designed for wired networks and the PBS (Pairwise Broadcast Synchronization) protocol that was designed for sensor networks, as none of them is able to provide the needed synchronization accuracy for our application on its own. The main goals of our new synchronization protocol are: to ensure the accuracy of local clocks up to a tenth of a microsecond and to provide an important energy saving. Our results obtained using NS-2 (Network Simulator) show that the performance of our solution (IEEE 1588-PBS) matches our application requirements with regard to the synchronization, with a significant improvement in energy saving.

## I. INTRODUCTION

Researches in the field of sensor networks show the variety and vastness of applications in which these types of systems are used. All these studies are done in order to improve the functionality of such communication systems. One of their main features is the large number (up to hundreds of elements) of sensors that must be distributed in different environments. This leads to the development of low-cost sensors [1], with limited processing capacity.

As such systems cannot be plugged into a continuous power supply and as the number of elements may be very high (which implies significant network traffic), a crucial problem is to minimize energy consumption [2]. In [3], a control solution for energy transmission for each node is proposed in order to extend the lifetime of the network without affecting the functionality of the system. Another proposal consists in making routing decisions in order to reduce the energy consumption. In [4] the proposal is to determine which node should be active (*awake* mode) at a time or not (*asleep* mode).

Depending on the application requirements, ensuring synchronous network functionality is currently a challenge. More specifically, the sensors collect information which they then send to the collector. The collector must receive the information from all sensors in a minimal timeframe: the time synchronization problem as described in [5]. This has lead, in traditional computer networks, to the design of many protocols used to maintain the synchronization of physical clocks. For instance, a protocol such as NTP (Network Time Protocol) is not a good choice for WSN, because of assumptions not valid in WSN [6].

The solutions proposed until now in literature, solutions which take into account these two aspects jointly (energy and synchronization), are not sufficient. In [7] the authors propose a clock synchronization protocol which is energy efficient, based on the estimation of the clock offset relative to a virtual clock.

The context of our study is related to the SACER project, funded by *Aerospace Valley* [8]. This project aims at designing a wireless sensor network of several hundred nodes located on aircraft wings, which will take pressure and temperature readings, during different flight test phases. The most important requirement in this application is the synchronization and precision of these readings in a very small temporal window of about several nanoseconds, so as to allow the correlation of the collected data and to improve the fluid dynamics modeling. So it was mandatory to ensure the synchronization performance without decreasing the lifetime of the network.

This paper describes our proposal based on the extension of IEEE1588-PBS (Pairwise Broadcast Synchronization [10]). Then we show its implementation and simulation and how its performance is compared to the one of IEEE 1588 [9].

The paper proposes a state of the art of existing research regarding the minimization of energy consumption and the problem of ensuring proper synchronization for Wireless Sensor Networks (Section II). Section III describes our proposed solution with its benefits supported by the results of the simulation. In Section IV we make a comparison between performance in terms of synchronization and power consumption between the IEEE 1588 standard and our protocol. Finally, Section V is dedicated to the conclusion and to future work.

## II. RELATED WORK

The problem of the synchronization of wired networks has been solved thanks to the development of successful protocols, in terms of their accuracy. Unfortunately, they are unsuitable for wireless sensor networks because the differences between wired and wireless networks are manifold. One of the most important differences is that sensors can suffer from power failure, which limits the use of existing technologies and communications protocols.

Our strategy is to exploit the performance of synchronization protocols for wired networks by adapting them for WSN. At the same time we consider that reducing energy consumption is essential to the extension of the lifetime of the network.

## *A. Synchronization Problems and Protocols*

The process of clock synchronization for distributed systems is to provide a common notion of time across the entire system. But the time of a software clock cannot be perfect. That is why we are interested in its accuracy, which can be calculated by analyzing parameters such as: the clock frequency, the clock offset, the skew and the drift of the clock [5]. Thus, NTP uses the method of Offset Delay Estimation [6], and performs the time synchronization of the central server with the UCT (Universal Coordinated Time). The disadvantage of this solution is that although it ensures high synchronization, it does it at the expense of message complexity.

More recently, the IEEE1588 standard [11] became the new reference in clock synchronization for industrial applications, as a result of its performance (accuracy to a hundredth of a microsecond (10ns)). But as wireless networks are limited in terms of size, power and complexity, most implementations of the IEEE 1588 are used on wired networks. On this matter, [12] presents experiments and a performance evaluation based on accurate time synchronization with IEEE1588 in WSNs. Results have shown that the synchronization between the master clock and the slave clocks of the network nodes is achieved with an accuracy to a tenth of a microsecond (100ns).

Several clock synchronization protocols have been proposed for wireless sensor networks, with better or worse performance. The Reference Broadcast Synchronization (RBS) [13] is the most representative protocol with a receiver-receiver scheme. By exploiting the broadcast property of the wireless communication medium, this protocol is able to achieve synchronization of a group of nodes that are in the communication range of a reference sender. Nodes which receive the broadcasted beacon will record the time of arrival and exchange this information with others. The precision of this protocol in 802.11with kernel time stamping is of 6.29 ± 6.45μs [13].

Timing-sync Protocol for Sensor Networks (TPSN) is an implementation of the sender-receiver synchronization method. This concept consists of two phases: the discovery and the synchronization. In [14] the authors implemented TPSN on Berkeley's Mica architecture and proposed a MAC layer time-stamping procedure, which is able to efficiently reduce the medium access time. The average error obtained is 16.9 μs.

The Pairwise Broadcast Synchronization (PBS) scheme proposed in [10] describes a new synchronization approach, called receiver-only synchronization (ROS) for a network-wide synchronization. The accuracy is similar to that obtained for RBS (tens of μs) on Berkeley Motes. PBS requires only $N_{PBS}$ =2N timing messages for each synchronization cycle where N represents the number of exchanged messages. Also, N does not depend on the number of network elements, which implies a great advantage in terms of energy savings.

## *B. Energy Problem*

One of the most important parameters for a WSN is its energy consumption, because of the limitations imposed upon it. For this reason existing studies propose different strategies to achieve this challenge.

The first proposal is to improve routing protocols, which can be divided into three major categories: data-centric, hierarchical and location-based [2]. Data-centric protocols show the advantage of eliminating many redundant transmissions; the hierarchical ones, through aggregation, can achieve energy savings, while the last category (location based protocols) is designed to provide some QoS (Quality of Service) capabilities along with the routing function [16].

On the other hand, solutions that rely on energy-efficient MAC protocols have several advantages [17]. Choosing a TDMA medium access is attractive because it eliminates the risk of collisions and the network becomes self-organized due to slot assignment and synchronization. Moreover, combining the advantages of TDMA and those of cluster topology increase energy efficiency [18].

As we can see, satisfying the need for synchronization and energy saving is not an easy task because these criteria are in opposition in terms of performance. More specifically, to ensure proper synchronization, the network will consume a significant amount of energy.

Researches for this purpose are quite recent [19]. In [20] and [21] the authors propose a new scheme of clock synchronization in order to optimize energy consumption and compare the results with performance of the TPSN protocol. To meet the need of synchronizing mobile nodes in a network [22] shows a hybrid solution that integrates the RBS and TPSN protocols.

We observed for this problem that PBS is the only synchronization protocol which, because of its *modus operandi*, is able to achieve energy savings as well.

## *C. Key Observations*

This overview of major existing solutions and protocols will guide us in the definition a new solution to meet the needs of our application. TABLE I summarizes a comparison between these existing solutions, based on various parameters, where I, N represents the number of exchanged packets in a synchronization cycle and L the number of network nodes.

Several key observations are found here to justify the use of our proposal in this wireless sensors application. So as to eliminate the risk of interference due to proximity of the components we made a choice for the TDMA medium access. As we noted in various studies in this field, choosing a hierarchical routing solution implies energy conservation and allows us to take advantage of the benefits of the aggregation method. Furthermore, if the data are continuously transmitted to the collector, it is more efficient to use a hierarchical structure. In the next section we present in detail the architecture of our network.

By analyzing TABLE I, it may be seen that existing synchronization protocols have either a very good synchronization accuracy with a high energy consumption or vice versa. Our idea is to take advantage of the performance, in terms of synchronization accuracy, of the IEEE1588 standard and then of the energy saving capability of the PBS protocol.

TABLE I. CLASSIFICATION ON SYNCHRONIZATION AND ENERGY ISSUES

| **Protocol** | **Synchronization issues** | | | | |
|---|---|---|---|---|---|
| | ***Scheme used*** | ***Number of messages*** | ***Time clock precision*** | ***Implementation level*** | ***Power Consumption*** |
| IEEE 1588[11] | Single hop | 4*N*L | 200 ns | App/ Physical | High |
| RBS[13] | Receiver-to-Receiver | $N*L^2$ | 29.1 µs | App | High |
| TPSN[14] | Sender-to-Receiver | 2*N*L | 16.9 µs | MAC | Average |
| FTSP[15] | Sender-to-Receiver | N*L | 1.7µs | MAC | Low |
| PBS[10] | Single et Multi hop | 2*N | 29.1 µs | App | Low |

By taking into account all the constraints of our system and thus choosing appropriate communication technologies, we have devised a new solution which is performant both in terms of accuracy and energy consumption.

## III. OUR PROPOSED SOLUTION

By analyzing how each of the aforementioned protocols is designed and considering our application requirements we concluded that none of the existing solutions is suitable for our system. So we envisioned the development of a new synchronization protocol.

### A. *Network Model Organization*

As mentioned earlier, our network is composed of a large number of items (64 active elements on each wing) that must communicate in flight, in order to centralize the collected information to the aircraft cabin (Fig. 1).

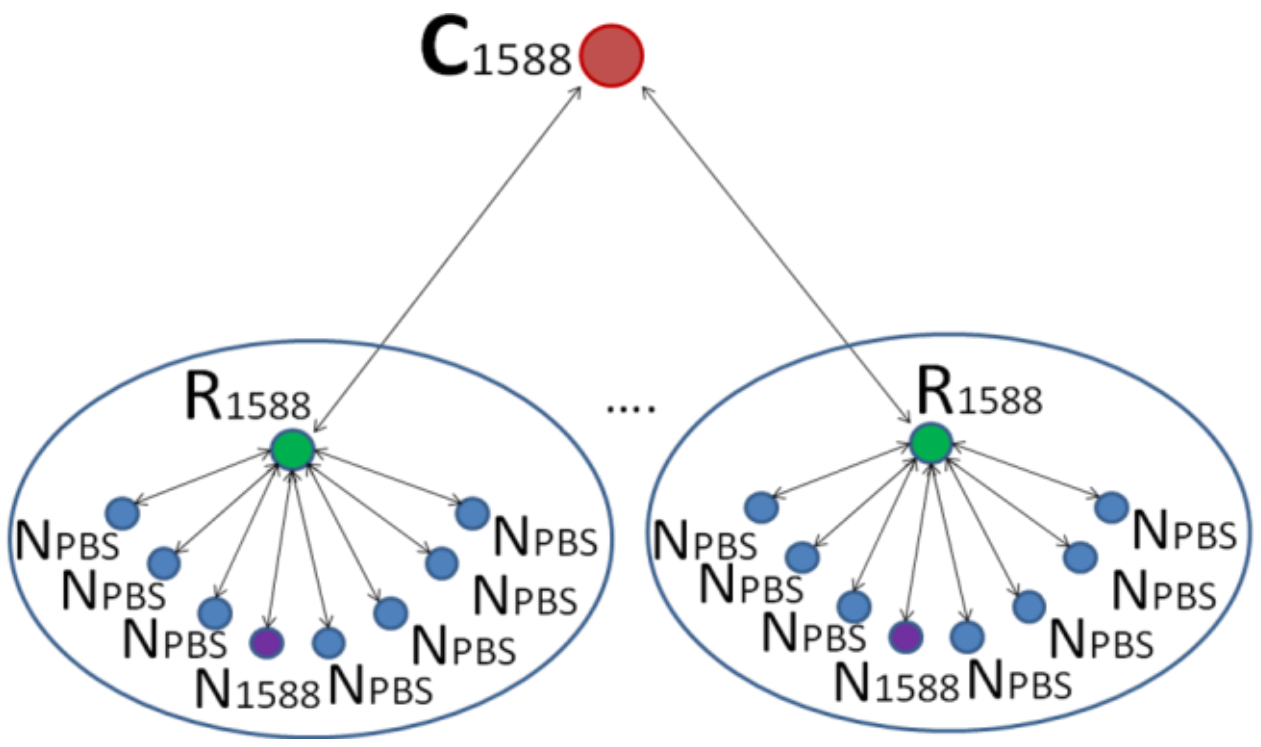

Figure 1. Network organization

Thus we propose a hierarchical organization solution which is composed of three distinct categories of elements (the concentrator, routers and nodes) in Fig. 1. The concentrator is designed to gather all information from routers and transmit them to the cockpit. Further, each one of the eight routers collects information from its nodes (which are eight in number as well).

### B. *The IEEE1588-PBS Hybrid Protocol*

In this sub-section we present our energy-efficient extension of the clock synchronization protocol IEEE 1588. In this approach, a group of nodes synchronizes by overhearing the timing messages of a slave-master pair, according to the PBS protocol principles. The way it works is explained in the following example of a group reduced to only 3 nodes. In such a group, the master (M) and the slave (S) nodes will synchronize using the IEEE1588 standard; the third node (X) will use the already existing synchronization (with high accuracy) between master and slave. This is done as in PBS with the assumption that node X is in the communication range of nodes M and S. Just by listening to the transmissions, X will synchronize its internal clock with that of node M with the help of the received messages.

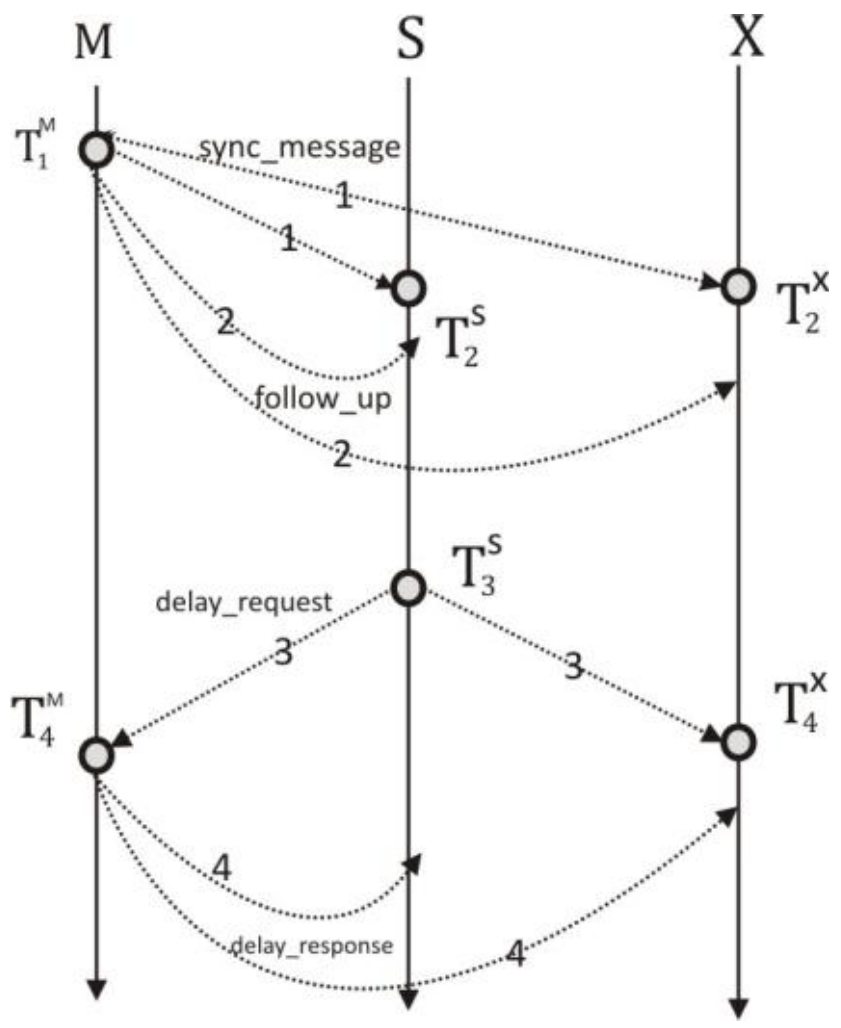

Figure 2. Exchanging messages in 1588-PBS protocol

Fig.2 shows how the protocol works for a 3 nodes group. The algorithm starts at node M by broadcasting a *sync_message* (1) to the slave and then storing at the physical layer the sent time-stamp ($T_1^M$). Afterwards, M broadcasts the *follow_up* message (2), containing the stored time-stamp. The first synchronization packet arrives at S at $T_2^S$ and at the PBS node (X) at $T_2^X$. Both nodes also store the receive time-stamps at the physical layer. In the next stage, at the slave's local time $T_3^S$, S broadcasts the request message 3 (*delay_request*). This is received by M at $T_4^M$ and also by X at $T_4^X$. M answers with “delay_reponse” (message 4), which contains the time-stamp $T_4^M$.

At the end of the synchronization session, X has received $T_1^M$, $T_2^X$, $T_4^M$, $T_4^X$, contained in received broadcasted messages. Let $\Delta_{XM}$ be the clock offset between X and M and $d_{XM}$ the propagation delay between X and M. We assume that the messages sent by S arrive at the master and at the PBS node at the same time (which implies that $d_{SM} - d_{SX} = 0$). So, the offset and the propagation delay between the nodes M and X become:

$$\Delta_{XM} = T_4^X - T_4^M \quad (1)$$

$$d_{XM} = T_2^X - T_1^M - \Delta_{XM} \quad (2)$$

According to Fig.2, the calculations performed to synchronize the node (X) of the network are given by:

$$T_2^X = T_1^X - \Delta_{XM} + d_{XM} \quad (3)$$

$$T_4^M = T_3^S - \Delta_{SM} + d_{SM} \quad (4)$$

$$T_4^X = T_3^X - \Delta_{SX} + d_{SX} \quad (5)$$

$$\Delta_{XM} = \Delta_{XS} + \Delta_{SM} \quad (6)$$

$$T_4^M - T_4^X = \Delta_{XM} + d_{SM} - d_{SX} \quad (7)$$

In the following section we present our results in terms of timing, number of messages and energy consumption, obtained by implementing our solution in the NS-2 Simulator.

## IV. Performance Evaluation and Comparison Study

This section provides a detailed quantitative analysis comparing the performance of our IEEE1588-PBS protocol and the standard IEEE1588.

The criteria taken into account in this evaluation are:

- the accuracy of synchronization,
- the power consumption in the nodes,
- the number of synchronization messages.

### A. *Performance Analysis*

To check and validate the behavior and the performance of the proposed solution, the essential settings needed in our evaluation have to be given first.

Thus, we started the synchronization mechanism in the first-level (concentrator-routers) at 0.3s and in the second level (routers-nodes) after 7s. In this way, nodes change their internal clock by taking as reference the clock of the router which is already synchronized.

To obtain precise time synchronization in WSN, it is necessary that the clock of the processor be controlled by a TCXO (Temperature Compensated Crystal Oscillator) at 37.5MHz [12], which has a 1.5PPM frequency tolerance, in order to reduce the drift rate. For our simulation, the internal clock of each node in the network was implemented with a drift of 1.5 microseconds per second. To achieve high performance and then to compare our protocol with the IEEE1588 standard, it was also decided to time-stamp the messages at the Physical layer. These information are then used in the synchronization scheme at the Application level.

Another important aspect is the hierarchical organization of our system, which is achieved through cooperation between the MAC, Routing and Transport levels.

### B. *Synchronization accuracy*

Our simulations were performed for the complete network (one concentrator, 8 routers and 64 nodes). Due to the hierarchical organization, the behaviour for a subgroup of one concentrator (C), one router (R) and eight nodes (N) is similar to that of any other subgroup. For this reason and because we are limited in number of pages, the graphical results only concern one such subgroup.

It is shown in Fig. 3 that the accuracy of the synchronization between the concentrator and the routers is around 1 to 2 hundred nanoseconds. We note that throughout the simulation the network elements remain synchronized at a high level (50-250ns).

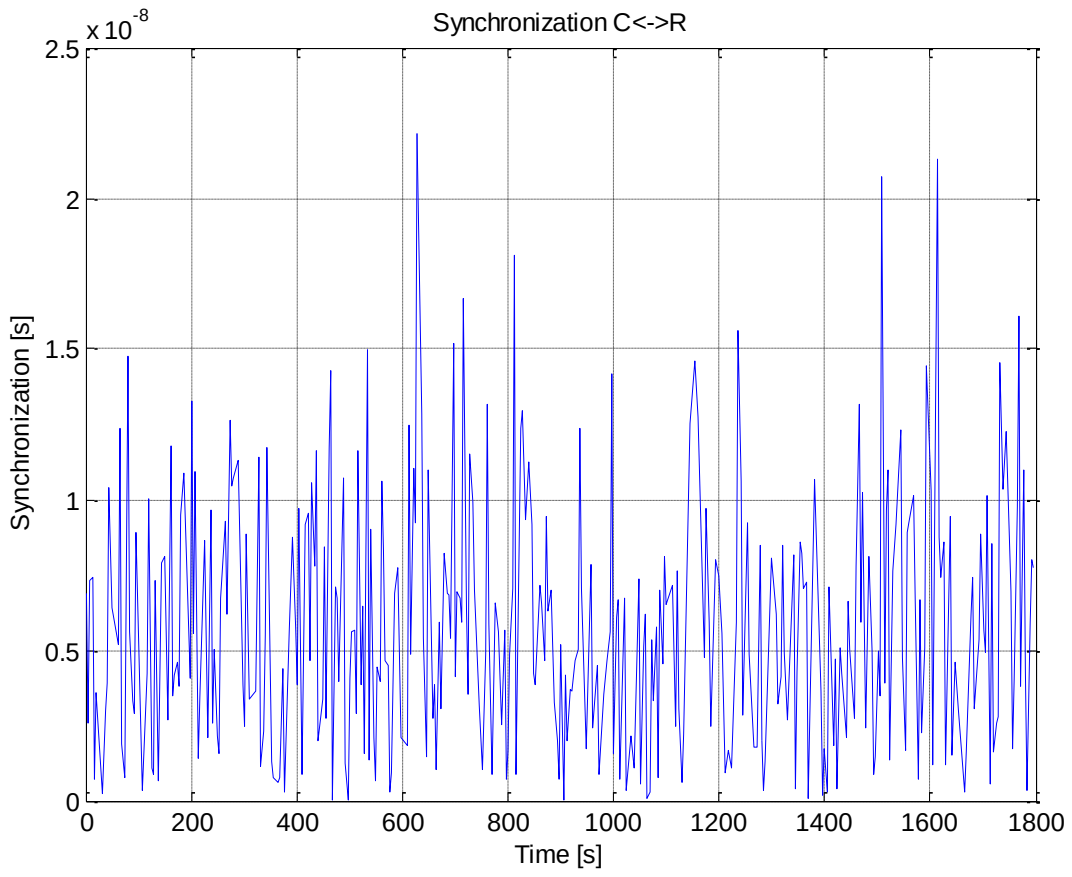

Figure 3. IEEE1588-PBS C<->R synchronization

Please remember that the IEEE1588 protocol results are quite the same (100-200ns), which is normal as the synchronization algorithm is the same. At the Router-Nodes level (Fig. 4), we notice that for the Router-Node1588 pair the synchronization has an order of magnitude of hundreds of nanoseconds and for the Router-NodePBS pairs it varies between 1µs and 46 µs.

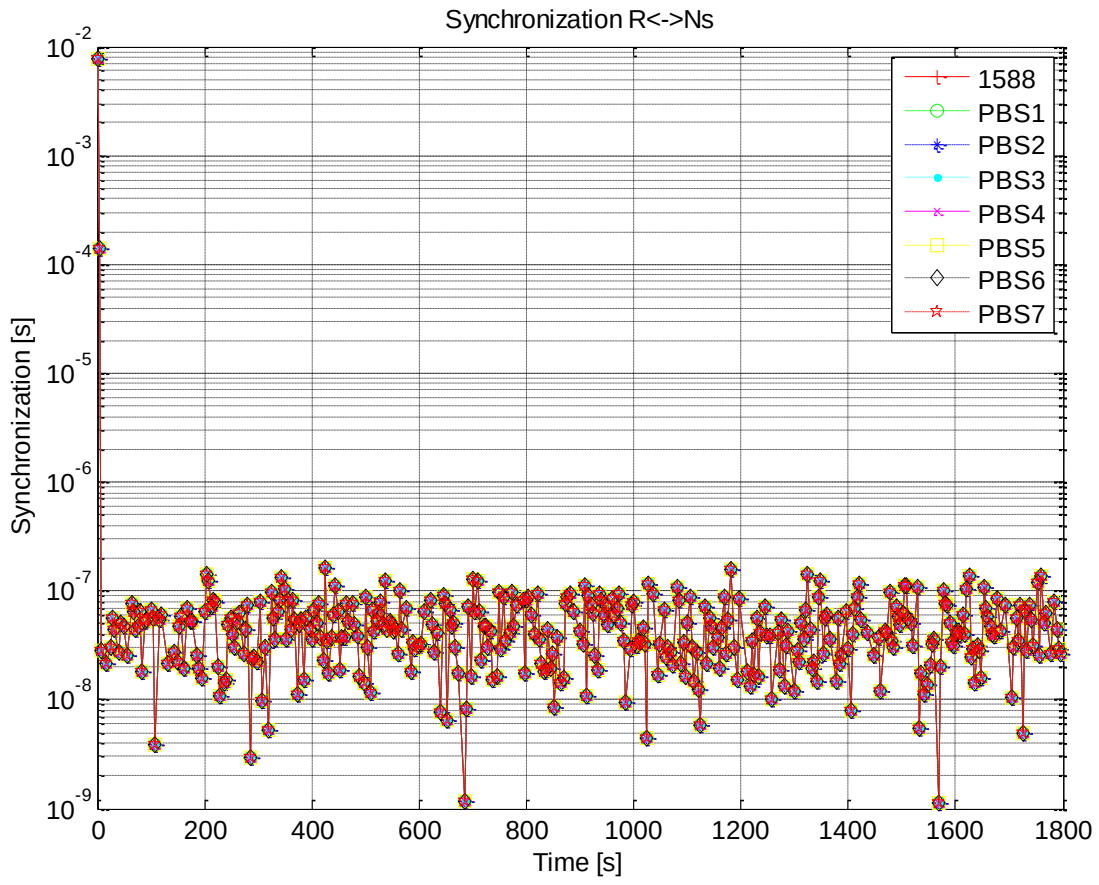

Figure 4. IEEE1588-PBS Rs<->Ns synchronization

The results for the IEEE1588 standard in the R<->Ns level are comparable with the accuracy of the Router-Node1588 pair in our solution (100-250ns).

We specify that the network wide synchronization is achieved in about 20 seconds after the beginning of the simulation.

So it is clearly shown by these results that the achieved accuracy is very good, close to the best ones found in literature.

### C. *Energy evaluation and number of messages*

In this section we are interested in the analysis of the energy consumption of our system, but especially the consumption in the nodes. We decided to use the energy consumption parameters of a Berkeley mote [12] in order to simulate real life conditions. In our network the energy consumption of sending a message is evaluated by NS-2 as 7mW and that of receiving a message as 4.5mW. Total initial energy is 2700J, which matches a CR2032 cell (3V, 230mA).

What interests us is the difference between energy consumption for the two protocols. Fig. 5 shows that the difference in energy consumption between a PBS node and a 1588 node is in the order of 15.07J. In other words, a PBS node consumes 84% less than a 1588 node. This is directly related to the number of messages required to achieve synchronization.

Thus we present in TABLE II the number of synchronization messages for the two analyzed protocols:

TABLE II. SYNCHRONIZATION MESSAGES

| **Protocols** | **Number of messages** | |
|---|---|---|
| | ***Level C<->Rs*** | ***Level Rs<->Ns*** |
| IEEE1588-PBS | 120messages/cycle | 24messages/cycle |
| IEEE1588 | 120messages/cycle | 1024messages/cycle |

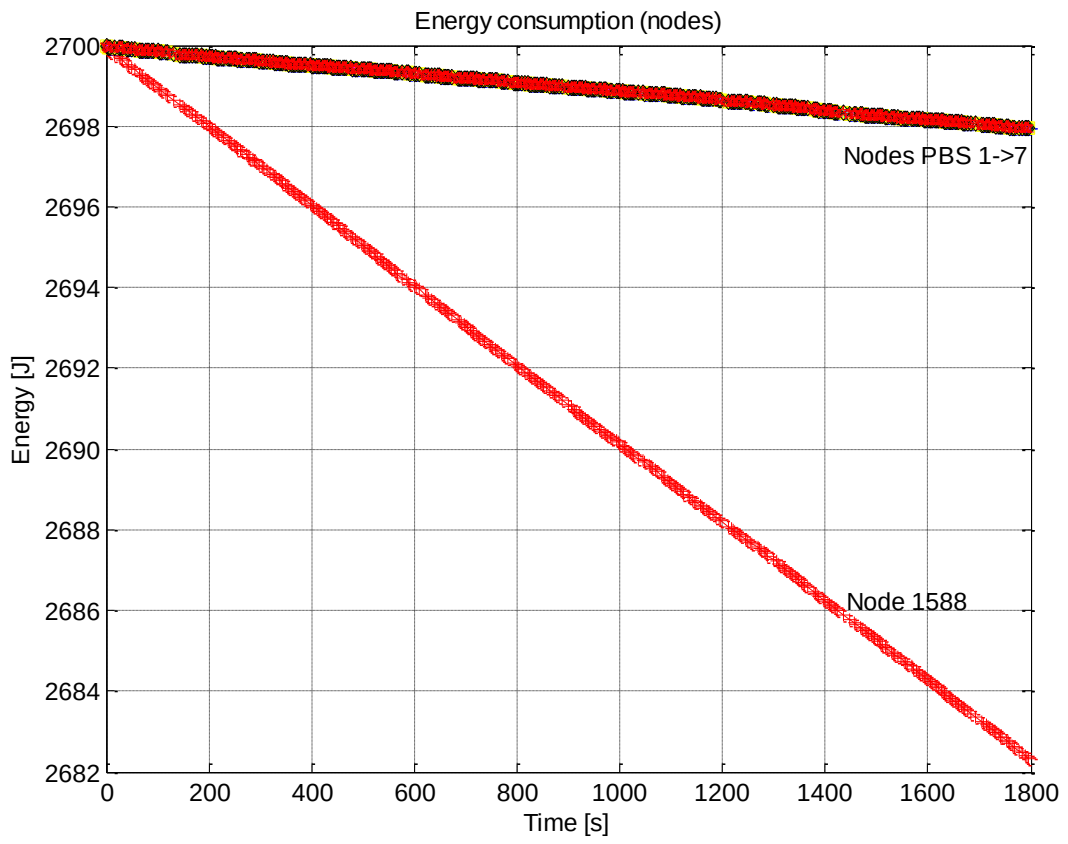

Figure 5. Energy Consumption for PBS and 1588 Nodes

This analysis allows us to see why our solution is more suitable for a wireless sensor network.

We proved that with good synchronization accuracy our solution achieves at the same time satisfactory energy savings.

## V. CONCLUSION AND FUTURE WORK

The protocol extension we proposed based on mixing the IEEE1588 and the PBS protocols for a hierarchical structure was simulated using NS-2. The obtained results show that this solution is able to obtain synchronization accuracy at a level of a tenth of microsecond (100ns) with efficient energy savings at the same time.

As the scalability of this solution is very important in the very next future, the number of nodes has to be extended to study the impact of the hierarchy. We also plan to use directional antennas to reduce consumption and interference problems. Another improvement may be done using 1 TDMA per subgroup and not only one for the entire WSN.